\documentclass[showpacs,preprintnumbers,amsmath,amssymb,12pt,floatfix,epsfig]{revtex4}
\usepackage{graphicx}
\begin{document}
\draft
\title{Roles of Direct Break-up Reaction on
Neutrino Scattering off $^{12}$C near Nucleon Threshold Region}
\author{K. S. Kim$^{1)}$\footnote{kyungsik@hau.ac.kr},
Myung-Ki Cheoun${^{2)}}$\footnote{Corresponding author :
cheoun@ssu.ac.kr}}
\address{1)School of Liberal Arts and Science, Korea Aerospace University,
Koyang 412-791, Korea  \\
2)Department of Physics, Soongsil University, Seoul 156-743, Korea
}

\begin{abstract}
Neutrino (antineutrino) scattering off $^{12}$C is one of various
important key reactions for $\nu$-process in the nucleosysnthesis
of light nuclei. Most of neutrino-nucleus scattering are
considered through indirect processes within the energy range from
a few to tens of MeV. Target nuclei are excited by incident
neutrino (antineutrino) through various transitions, and
subsequently decay into other nuclei with emitting particles. But,
direct processes are also feasible, in which incident neutrino
(antineutrino) strips directly one nucleon from target nuclei.
Consequently, direct processes may affect abundances of $^{11}$C
and $^{11}$B additionally to indirect processes. We investigate
direct neutrino (antineutrino) quasi-elastic scattering off
$^{12}$C around the energy region liberating one nucleon and
discuss implications of direct processes in the nucleosynthesis.
The direct processes might be comparable to the indirect processes
if the final state interaction is taken into account.
\end{abstract}
\narrowtext \maketitle

Neutrino ($\nu$) (antineutrino (${\bar \nu}$)) scattering with a
complex nucleus plays important roles of studying $\nu$
properties, such as $\nu$ oscillation and masses, as well as
nuclear structure probed by weak interaction
\cite{Woosley90,yoshida08,Suzuki06}. Hence, a lot of interests for
neutrino (antineutrino)-nucleus ($\nu ({\bar \nu} ) - A$)
scattering have been increased to the nuclear astrophysics, for
instance, $\nu$-process in the formation of a core collapsing
supernova, because cross sections for the $\nu ({\bar \nu} ) - A$
scattering are one of the most important input data for a network
calculation estimating light nuclei abundance like $^{7}$Li and
$^{11}$B. The abundance ratio turns out to be sensitive to the
$\nu$ oscillation parameters, mass hierarchy, and mixing angle
$\theta_{13}$ \cite{yoshida08}. Incident $\nu ({\bar \nu})$
energies exploited in these calculations
\cite{Woosley90,yoshida08} are focused on the energy range from a
few to a few tens of MeV, because relevant $\nu ({\bar \nu})$
energy spectra emitted from a proto-neutron star are presumed to
be mostly peaked around the energy region.

Most of calculations for the $\nu ({\bar \nu}) - A $ scattering
are performed by considering indirect processes. Incident $\nu
({\bar \nu})$ leads target nuclei to some excited states through
various transitions {\it i.e.}, super allowed Fermi ($J^{\pi} =
0^+)$, allowed Gamow Teller ($J^{\pi} = 1^+)$, spin dipole
($J^{\pi} = 0^- , 1^-, 2^-)$, and other higher multipole
transitions. The excited nuclei subsequently decays into other
nuclei with emitting particles such as proton, neutron, alpha,
$\gamma$, and so on \cite{Suzuki06,yoshida08}.

Since weak interaction is mediated by $Z^0$ and $W^{\pm}$ bosons,
there are two kinds of reactions, charge current (CC) and neutral
current (NC) reactions. In the NC reaction, the incident $\nu
({\bar \nu})$ excites target nuclei, and then the excited target
nuclei are subsequently decayed into other nuclei by emitting some
particles incoherently,
\begin{equation}
 A ( \nu ({\bar \nu}) , \nu^{'} ({\bar \nu}^{'} )) A^*
~,~ ~~A^* \rightarrow B + \mbox{outgoing particles}.
\end{equation}
$^{12}$C$( \nu , \nu^{'} ) ^{12}$C$^* \rightarrow ^{11}$B + p (or
$ ^{11}$C + n) reaction is one of the indirect processes for $ \nu
- ^{12}$C scattering. But, the direct knocked-out processes are
also possible \cite{Kosmas07}, in which a nucleon inside nuclei is
stripped from target nuclei without any excitation of target
nuclei. For instance, $ ^{12}$C$( \nu , \nu^{'} p ) ^{11}$B or
$^{12}$C$( \nu , \nu^{'} n ) ^{11}$C reactions should be
differentiated from the indirect processes.

Meanwhile, in the CC reaction, we have another direct processes,
$^{12}$C$( \nu_e , e^- p ) ^{11}$C and $^{12}$C$( {\bar \nu}_e ,
e^+ n ) ^{11}$B, in addition to the indirect processes
\begin{equation}
 A ( \nu_l ({\bar \nu}_{l}) , l ({\bar l}) ) B^*
~,~ ~~B^* \rightarrow C + \mbox{outgoing~particles} ~,
\end{equation}
which are $ ^{12}$C$( \nu_e , e^- ) \rightarrow ^{12}$N$^*
\rightarrow ^{11}$C + $p$ and $^{12}$C$( {\bar \nu}_e , e^+ )
\rightarrow ^{12}$B$^* \rightarrow ^{11}$B$ + n$. Therefore, it is
possible for the direct processes to influence the abundances of
redundant nuclei in the network calculation initiated from
$^{12}$C by the $\nu$-process \cite{yoshida08}.

A few experimental data for the $\nu ({\bar \nu}) - ^{12}$C
reaction through the indirect processes have been reported as flux
averaged out total cross sections since 1990. Detailed references
are summarized at Refs. \cite{Suzuki06,Kolbe96}. The data for
inclusive reaction such as $^{12}$C$( \nu_e , e^- ) ^{12}$N$^*$
show about $4.3 \sim 5.7$, while the data for exclusive reaction
like $^{12}$C$( \nu_e , e^- ) ^{12}$N$_{g.s.}$ are restricted to
$8.9 \sim 10.5$ in the $10^{-42}$ cm$^2$ unit. All these data are
measured from accelerated-based data. Future $\nu$ factory for
intense and pure $\nu$ ($\bar \nu$) beam, so called as beta beam
\cite{Volpe07}, could yield more fruitful data for the $\nu (
{\bar \nu}) - A$ scattering.

Many theoretical calculations
\cite{Kolbe95,Kosmas07,Kolbe96,Volpe00,Suzuki06,Ring08} have been
reported for the $ \nu ( {\bar \nu}) - ^{12}$C scattering since
the pioneering work on weak interactions on $^{12}$C by J. S.
Cornnell {\it et al.} \cite{Con72}. Of course, all of theoretical
results only assume indirect processes. Results of most shell
model (SM) calculations \cite{Suzuki06,Volpe00} converge more or
less to the experimental data although they depend on the particle
model space and the given Hamiltonian. But the results by the
random phase approximation (RPA) (or Continuum RPA)
\cite{Kolbe96,Volpe00,Kolbe95} and Quasi-particle RPA (QRPA)
\cite{Kosmas07,Volpe00,Ring08} calculations overestimated the data
by a factor of $ 4 \sim 5$, in specific, for the exclusive
reaction. Since the energy weighted sum rule is satisfied in the
RPA approach \cite{Volpe00}, these differences seem to be
inescapable and are claimed to come from the small model space
basis used in the SM calculation \cite{Suzuki06,Volpe00}.

In this report, we presume that the $\nu ({\bar \nu}) - ^{12}$C
scattering can be also proceeded via direct processes, and present
results for the direct CC reactions, $^{12}$C$( \nu_e , e^- )$ and
$^{12}$C$( {\bar \nu}_e , e^+ )$, and results for the direct NC
reaction, $^{12}$C$( \nu , \nu^{'})$, where ground states of
$^{11}$C and $^{11}$B are taken as final nuclei with all summation
of possible knocked-out nucleon states. In the Born approximation,
the CC reactions, $^{12}$C$( \nu_e , e^- )$ and $^{12}$C$( {\bar
\nu}_e , e^+ )$, are calculated by integrating the kinetic energy
and all the possible states of knocked-out nucleon on the direct
processes, $^{12}$C$( \nu_e , e^- p)$ and $^{12}$C$( {\bar \nu}_e
, e^+ n )$, respectively. Likewise, the NC reactions, $^{12}$C$(
\nu , \nu^{'} )$ and $^{12}$C$( {\bar \nu} , {\bar \nu}^{'} )$,
are given by the integration of the kinetic energy and all the
possible states of outgoing nucleon on the direct processes,
$^{12}$C$( \nu , \nu^{'} N )$ and $^{12}$C$( {\bar \nu} , {\bar
\nu}^{'} N)$, respectively. Since we consider direct processes,
the excited states of final nuclei are not considered.

The direct processes considered here just correspond to a low
energy tail of quasi-elastic scattering peak. And to calculate
this procedure we use the distorted wave Born approximation (DWBA)
formalism which has been successfully applied to the quasi-elastic
electron scattering for a long time.

Since the framework of the DWBA is focused on a nucleon inside
nuclei, main ingredients are wave functions of bound and continuum
nucleons, and a weak transition current operator. Detailed
descriptions are given in our previous papers
\cite{kimplb,kimjpg07,kim1,kim2}, which satisfactorily described
the quasi-elastic $\nu -A$ \cite{kimplb,kimjpg07} as well as the
electron-nucleus scattering \cite{kim1,kim2}.

For obtaining the nucleon bound state wave functions, the Dirac
equation is solved in the presence of the strong vector and scalar
potentials based on $\sigma-\omega$ model \cite{horo}. The wave
functions of the continuum nucleons are the solution of the Dirac
equation with a relativistic phenomenological optical potential
generated by Ohio State University group \cite{clark}.

We choose the nucleus fixed frame where target nucleus is seated
at the origin of the coordinate system. Four-momenta of incident
and outgoing $\nu ({\bar \nu})$ are labelled $p_i^{\mu}=(E_i, {\bf
p}_i)$, $p_f^{\mu}=(E_f, {\bf p}_f)$. $p_A^{\mu}$,
$p_{A-1}^{\mu}$, and $p^{\mu}$ represent four-momenta of target
nucleus, residual nucleus, and final nucleon, respectively. In the
laboratory frame, the differential cross section is given by the
contraction between the lepton tensor and the hadron tensor
\cite{kimplb}
\begin{eqnarray}
{\frac {d\sigma} {dT_p}} = 4\pi^2{\frac {M_N M_{A-1}} {(2\pi)^3
M_A}} \int \sin \theta_l d\theta_l \int \sin \theta_p d\theta_p p
f^{-1}_{rec} \sigma^Z_M [v_L R_L + v_T R_T + h v'_T R'_T ],
\label{cs}
\end{eqnarray}
where $M_N$ is the nucleon mass, $\theta_l$ denotes the scattering
angle of the lepton, and $h=-1$ $(h=+1)$ corresponds to the
helicity of the incident $\nu ({\bar \nu})$. $\theta_p$ and $T_p$
represent the polar angle and the kinetic energy of the
knocked-out nucleons, respectively. For the NC reaction,
$\sigma^Z_M$ is defined by
\begin{equation}
\sigma^Z_M = \left ( {\frac {G_F \cos (\theta_l/2) E_f M_Z^2}
{{\sqrt 2} \pi (Q^2 + M^2_Z)}} \right ),
\end{equation}
and for the CC reaction,
\begin{equation}
\sigma^{W^\pm}_M = \sqrt{1 - {\frac {M^2_l} {E_f}}} \left ( {\frac
{G_F \cos (\theta_C) E_f M_W^2} {2\pi (Q^2 + M^2_W)}} \right )^2,
\end{equation}
where $G_F$ is the Fermi constant given by $G_F \simeq 1.16639
\times 10^{-11}$ MeV$^{-2}$, and $M_Z$ $(M_W)$ is the rest mass of
$Z$ ($W$)-boson. $\theta_C$ denotes the Cabibbo angle given by
$\cos^2 \theta_C \simeq 0.9749$. Detailed forms for recoil factor
$f_{rec}$, kinematical coefficients $v$, and the corresponding
response functions $R$ are given in Ref.\cite{kimplb}.

The nucleon current $J$ represents the Fourier transform of the
nucleon current density written as
\begin{equation}
J^{\mu}=\int {\bar \psi}_p {\hat {\bf J}}^{\mu} \psi_b e^{i{\bf
q}{\cdot}{\bf r}}d^3r,
\end{equation}
where ${\hat {\bf J}}^{\mu}$ is a free nucleon current operator,
and $\psi_{p}$ and $\psi_{b}$ are the wave functions of the
knocked-out and the bound state nucleons, respectively. Total
cross section is given as the integration of Eq.(3) to the kinetic
energy of the knocked-out nucleon:
\begin{equation}
\sigma = \int {\frac {d \sigma} {dT_p}} dT_p .
\end{equation}

In Figs. 1 - 3, we show total cross sections for the NC and CC
reactions by the direct processes in terms of the incident $\nu$
($\bar \nu$) energy, {\it i.e.}, $^{12}$C$( \nu , \nu')$, and
$^{12}$C$( \nu_e , e^- )$ and $^{12}$C$( {\bar \nu_e} , e^+ )$,
respectively. Since the threshold energy for liberating nucleon is
just the binding energy of the nucleon inside nuclei, our results
are presented from the averaged binding energy. Emitting muon in
the CC reaction is energetically forbidden on the energy region
considered here.

Our results are separately presented with and without an optical
potential, which is introduced to take final state interaction
(FSI) of outgoing nucleon with residual nuclei into account. With
the FSI, cross sections are generally reduced by a factor of 2
compared with those without the FSI. This reduction also appears
on other calculations \cite{kimjpg07,giusti2}. In specific, the
FSI affects in the whole energy region. Therefore, the FSI of
outgoing nucleon could be one of vital important ingredients even
in the indirect processes on the low energy region.

Results of the indirect processes symbolized as data points in the
all figures are taken from the SM calculation tabulated in Ref.
\cite{yoshida08}. They present two theoretical calculations based
on two different Hamiltonian, SFO and PSDMK2 \cite{Suzuki06}. No
remarkable difference between the two results can be seen in the
log scale cross sections.

Comparison of our results {\it i.e.}, direct processes including
the FSI, to those by the indirect processes reveals that the cross
sections of the direct processes are smaller by a factor of 2
$\sim$ 3 for $^{12}$C$( \nu , \nu' )$ and $^{12}$C$( {\bar \nu}_e
, e^+ )$, and by a factor of 3 $\sim$ 4 for $^{12}$C$( \nu_e , e^-
)$ rather than those of the indirect processes (see differences
between data points and solid curves at figures). It means that
the contributions of the direct processes to the abundance of
light nuclei could be small by a factor of 2 $\sim$ 4 compared to
those by the indirect processes.

However, it should be noted that most of calculations for the
indirect processes did not take into account of the FSI, and the
FSI due to the strong interaction of outgoing nucleons with
residual nuclei could lower the cross sections in the low energy
$\nu ({\bar \nu})$, even if outgoing particles emit from compound
nuclei. As shown in Figs. 1 - 3, the reduction of cross sections
by the optical potential at the nucleon threshold energy may
support such a conjecture. Therefore, the relatively small present
contribution of the direct processes could be comparable or even
larger to those of the indirect processes if the FSI effects could
be taken into account in the indirect processes.

In order to compare with forthcoming experimental data, we should
need to consider flux averaged (folded) cross section, although
there are still no data for the direct processes. It needs to know
the neutrino energy spectrum which inevitably depends on a given
temperature, just like the Fermi distribution usually adopted in
most calculation. Detailed studies of cross sections by direct
processes to the given temperature and their effects on the
abundances in the network calculation of nucleosynthesis are in
progress.

\section*{Acknowledgements}
This work was supported by the Soongsil University Research Fund.
Cheoun is greatly indebted for discussions with T. Kajino, T.
Yoshida and T. Suzuki, and the hospitality of NAOJ, Japan.

\newpage

\begin{figure}
\includegraphics[width=0.85\linewidth]{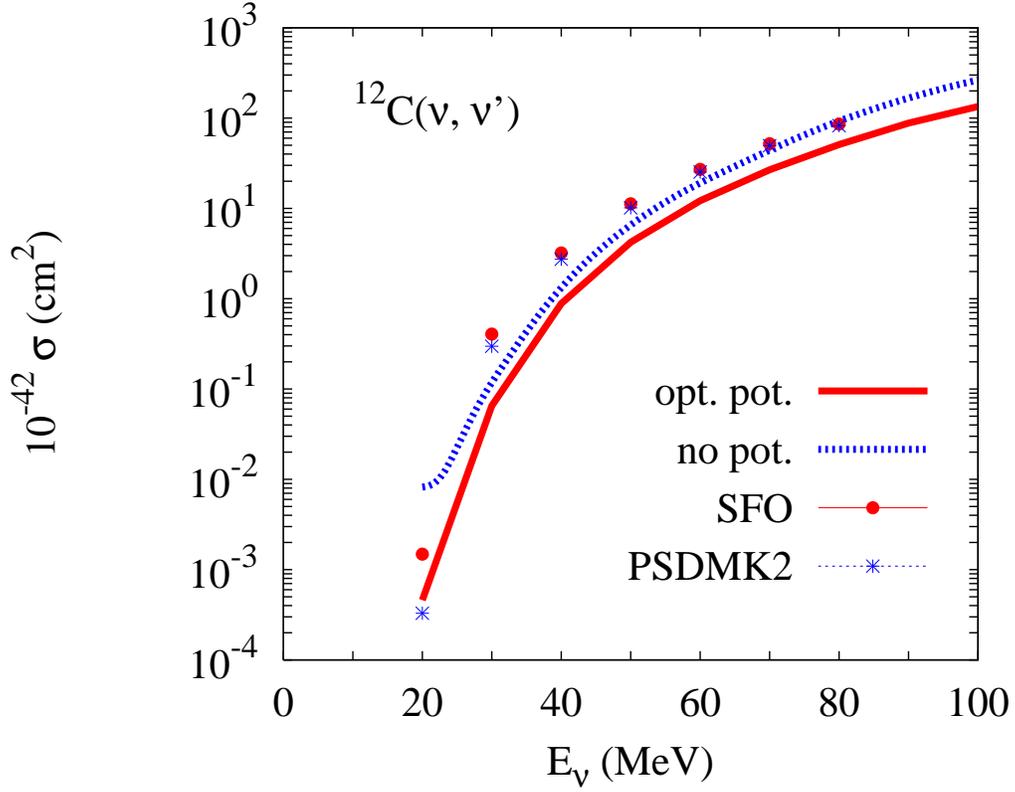}
\caption{(Color online) NC reaction for ${ \nu}$ by direct
process, $^{12}$C$( \nu , \nu' )$, obtained by integrating the
kinetic energy and summing all possible knocked-out nucleon states
for $^{12}$C$( \nu , \nu' N )$ reaction \cite{kimplb}. Data points
for indirect processes, which is a sum of two cross sections,
$^{12}$C$( \nu , \nu^{'} ) ^{12}$C$^* \rightarrow ^{11}$B + p and
$ ^{11}$C + n, come from the SM calculation \cite{yoshida08}. SFO
and PSDMK2 mean two different Hamiltonian exploited in the
calculation.} \label{fig1}
\end{figure}

\begin{figure}
\includegraphics[width=0.85\linewidth]{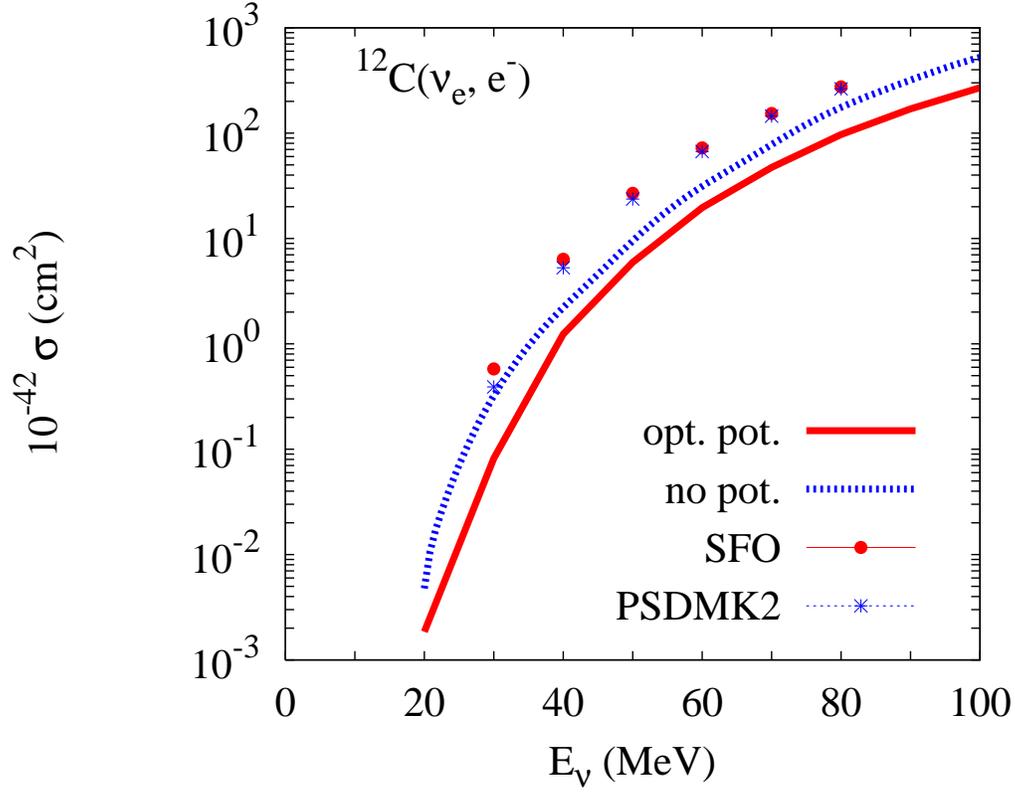}
\caption{(Color online) CC reaction for $\nu_e$ by direct process,
$^{12}$C$( \nu_e , e^- )$, obtained by integrating the kinetic
energy and summing all possible knocked-out proton states in the
reaction, $^{12}$C$( \nu_e , e^- p)$. Data points for indirect
processes come from the SM calculation for $ ^{12}$C$( \nu_e , e^-
) \rightarrow ^{12}$N$^* \rightarrow ^{11}$C + $p$
\cite{yoshida08}. Others are same as Fig.1.} \label{fig2}
\end{figure}

\begin{figure}
\includegraphics[width=0.85\linewidth]{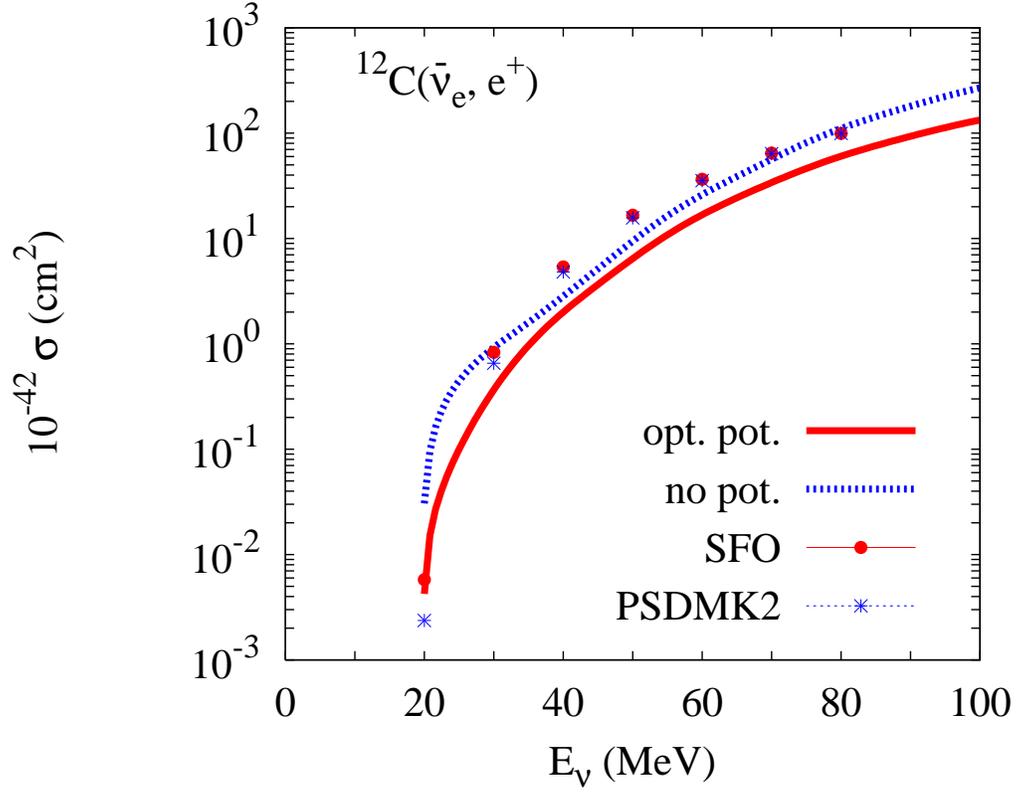}
\caption{(Color online) CC reaction for ${\bar \nu}_e$ by direct
process, $^{12}$C$( {\bar \nu}_e , e^+ )$, obtained by integrating
the kinetic energy and summing all possible bound neutron states
in the reaction, $^{12}$C$( {\bar \nu}_e , e^+ n)$. Data points
for indirect processes come from the SM calculation for $^{12}$C$(
{\bar \nu}_e , e^+ ) \rightarrow ^{12}$B$^* \rightarrow ^{11}$B$ +
n$ \cite{yoshida08}. Others are same as Fig.1.} \label{fig3}
\end{figure}

\end{document}